\documentclass[review]{elsarticle}

\usepackage{lineno,hyperref}
\modulolinenumbers[5]
 
\journal{Computational Statistics \& Data Analysis}
\usepackage[utf8]{inputenx}
\usepackage{subfig}
\usepackage{appendix}
\usepackage{pdfpages}
\usepackage{slashed}
\usepackage{float}
\usepackage{amsmath}%
\usepackage{amsfonts}%
\usepackage{cancel}
\usepackage{amssymb}%
\usepackage{multirow} %
\usepackage{xcolor}

\begin{document}

\begin{frontmatter}

\title{{An improved method for the estimation of the Gumbel distribution parameters\tnoteref{mytitlenote}}
\tnotetext[mytitlenote]{poner despues}}

\author[mymainaddress]{Rub\'en G\'omez Gonz\'alez\corref{mycorrespondingauthor}}
\cortext[mycorrespondingauthor]{Corresponding author}
\ead{ruben@unex.es}
\author[mysecondaryaddress,insttres]{M. Isabel Parra}
\ead{mipa@unex.es}
\author[mymainaddress,instuno]{Francisco Javier Acero}
\ead{fjacero@unex.es}
\author[mysecondaryaddress,instdos]{Jacinto Mart\'in}
\ead{jrmartin@unex.es}

\address[mymainaddress]{Departamento de F\'isica, Universidad de Extremadura, Avenida de Elvas,  
	06006 Badajoz, Spain}
\address[mysecondaryaddress]{Departamento de Matem\'aticas, Universidad de Extremadura, Avenida de Elvas, 06006 Badajoz, Spain}
\address[insttres]{Instituto de Investigaci\'on de Matem\'aticas de la Universidad de Extremadura (IMUEX), Universidad de Extremadura,  Avenida de Elvas, 06006 Badajoz, Spain}
\address[instuno]{Instituto Universitario de Investigaci\'on del Agua, Cambio Clim\'atico y Sostenibilidad (IACYS), Universidad de Extremadura,  Avenida de Elvas, 06006 Badajoz, Spain}
\address[instdos]{Instituto Universitario de Computaci\'on Cient\'ifica Avanzada de Extremadura (ICCAEX), Universidad de Extremadura,  Avenida de Elvas, 06006 Badajoz, Spain}

\begin{abstract}
Usual estimation methods for the parameters of extreme values distribution employ only a few values, wasting a lot of information. More precisely, in the case of the Gumbel distribution, only the block maxima values are used. In this work, we propose a method to seize all the available information in order to increase the accuracy of the estimations. This intent can be achieved by taking advantage of the existing relationship between the parameters of the baseline distribution, which generates data from the full sample space, and the ones for the limit Gumbel distribution. In this way, an informative prior distribution can be obtained. Different statistical tests are used to compare the behaviour of our method with the standard one, showing that the proposed method performs well when dealing with very shortened available data.    
The empirical effectiveness of the approach is demonstrated through a simulation study and a case study. Reduction in the credible interval width and enhancement in parameter location show that the results with improved prior adapt to very shortened data better than standard method does.
\end{abstract}


\end{frontmatter}

\section{Introduction}

Extreme value theory (EVT) is a widely used statistical tool for modeling and forecasting an extreme value distribution of a given sample generated by a baseline distribution. EVT is employed in several scientific fields, for instance in climate change through studies of extreme events of temperature \cite{nogaj2006amplitude,Coelho2008,Acero2017}, precipitation \cite{Garcia2007,Re2009,Acero2011,Acero2011a,Acero2017b,Wi2015}, and solar climatology \cite{Ramos2007,Acero2017c,Acero2018a}, or engineering where it is taken into account to design, for example, modern buildings \cite{Castillo2004}. The extreme values depend upon the full sample space from which they have been drawn through its shape and size. Therefore, extremes variate according to the initial distribution and the sample size \cite{Gumbel2012}. 

The Gumbel distribution belongs to the Generalized Extreme Value distribution group (GEV) used to shape the maximum (or minimum) distribution of a sample space which can arise from various baseline distributions. By being able to describe extreme events of the normal or exponential type \cite{castillo1987estadistica}, the Gumbel distribution is greatly useful for modeling experimental and social data, the main two areas in which EVT is used. In order to estimate maximum data distribution, both frequentist and Bayesian approaches have been developed \cite{Smith1987a,Coles2003a}. However, the knowledge of physical constraints, the historical evidence of data behaviour, or previous assessments might be an extremely important matter for the adjustment of the data, particularly when it is not completely representative and further information is required. This fact leads to the use of Bayesian inference to address the extreme value estimation \cite{Bernardo1994}.

Very few attempts have been made to incorporate Bayesian methodology into extreme value analysis. Nevertheless, practical use of Bayesian estimation is often associated with difficulties to choose prior information and prior distribution for Gumbel parameters \cite{Kotz2000}. To fix this problem, several alternatives have been proposed,  either by focusing exclusively on the selection of the prior density \cite{Coles1996,Rostami2013} or by improving the algorithm for the estimation of the parameters \cite{Chen2000}. Even so, the lack of information still seems to be the weak point when referring to extreme value inference. In this paper, a known baseline distribution for the complete dataset is assumed and a relationship between its parameters and the parameters of the Gumbel distribution is established through a bayesian iterative procedure. Thus, the aim of this work is to obtain a more accurate model for the estimation of the Gumbel parameters by considering a highly informative prior based on the connection between the distribution parameters.  The use of the entire data, instead of the selected maximum data, provides an upgraded method which results to adapt quite well to very shortened available data. 

More precisely, a regression analysis is implemented to estimate a relation between the parameters of the initial distribution and the parameters of the extreme distribution for a given sample size. Once this connection is obtained, the next step is to 
obtain the posterior distribution of the parameters of the Gumbel distribution. To perform this task, we need to use a MCMC method, concretely a Metropolis Hastings algorithm. Therefore, the proposed method is a regression Metropolis-Hastings algorithm (RMH). Several statistical analysis are performed to test the validity of our method (RMH) and to check its enhancements in relation to the Standard Metropolis Hastings (SMH) algorithm.

The paper is organized as follows. In Section 2 the asymptotic model used to describe extremal behaviour is outlined. Specifically, Gumbel distribution and Bayesian inference for the estimation of its parameters are introduced. In Section 3, the regression analysis performed to improve MCMC procedure is proposed and implemented, and its coefficients are implemented into the generic model which is described in Section 4. The results of our analysis are presented in Section 5. Here, we highlight the advantages that the knowledge of the regression coefficients has in this extreme value context over the using of non-informative priors. For the sake of completeness, Section 6 gives a real example of application of the model. 

\section{The block maxima method in extreme value theory}

In extreme value theory, there are two fundamental ways to model the limit distribution: the block maxima (BM) method and the peaks-over-threshold (POT) method. We center our work on the former, which is a rather efficient method under usual practical conditions. BM method may be preferable than POT in several situations, specially when the only available info is block maxima, seasonal periodicity is presented, or the block periods appear naturally.

The BM approach consists of dividing the observation period into non-overlapping periods of equal size and look at the maximum observation in each period. Let $\widetilde{X}_1, \widetilde{X}_2, ..., \widetilde{X}_m$ be i.i.d. random variables with distribution function $F$. Given a fixed $k \in \mathbb{N}$, we define the block maxima 
\begin{equation}
X_i=\max_{(i-1)k < j \leq ik} \widetilde{X}_j, \ i=1,2,...,n. \end{equation}
Hence, the $m=k \times n$ observations are divided into $n$ blocks of size $k$.

The new random variables created follow, under certain domain of attraction conditions, approximately an extreme value distribution.

In order to estimate maximum data distribution, frequentist and Bayesian methods have been developed. Here we focus on the second procedure. More precisely, we model the limit distribution by the BM method and estimate the parameters of this distribution by employing Bayesian techniques.

\section{Bayesian estimation of the Gumbel distribution}
Many procedures have been suggested, optimized, or discarded to enhance Bayesian analysis of Gumbel distributed data. Some examples are the modeling of annual rainfall maximum intensities \cite{Vidal2013}, the estimation of the probability of exceedance of future flood discharge \cite{Lye1990}, or the forecasting of the extremes of the price distribution \cite{Rostami2013a}. Some of these works are focused on the construction of informative priors of the parameters for which data can provide little information.  Despite these previous efforts, it is well understood that some constraints to quantify qualitative knowledge always appear when referring to construct informative priors. For this reason, this paper focuses on Bayesian MCMC techniques based on MH algorithms to estimate the parameters using non-informative prior \cite{1996}. 

In order to make statistical inferences based on the Bayesian framework, after assuming a prior density for the parameters, $\pi(\boldsymbol{\theta})$, and combining this distribution with the information brought by the data which is quantified by the likelihood function, $L(\boldsymbol{\theta}|x)$, the posterior density function of the parameters can be determined as follows
\begin{equation}
\pi(\boldsymbol{\theta}|x) \propto L(\boldsymbol{\theta}|x) \pi(\boldsymbol{\theta}).
\end{equation}
The remaining of the inference process is fulfilled based on the obtained posterior distribution.

Given the random sample $\mathbf{x}=(x_1,...,x_n)$ from a $Gumbel(\mu,\sigma)$ distribution, with {\it{cdf}} and density function given by
\begin{equation}
F(x|\mu, \sigma)=\exp{\left( -\exp{\left(-\frac{x-\mu}{\sigma}\right)}\right)}
\end{equation}
and
\begin{equation}
f(x|\mu, \sigma)=\frac{1}{\sigma}\exp{\left( -\exp{\left(-\frac{x-\mu}{\sigma}\right)} - \frac{x-\mu}{\sigma} \right)}
\end{equation}
respectively, where $\mu \in \mathbb{R}$, $\sigma \in \mathbb{R}_{+}^{*}$ and $x$ can take any value in $\mathbb{R}$, the likelihood function for $\boldsymbol{\theta}=(\mu,\sigma)$ is
\begin{equation}\label{likehood}
L(\mu, \sigma|\mathbf{x})=\frac{1}{\sigma^n}\exp{\left( - \sum_{i=1}^n{\exp{\left(-\frac{x_i-\mu}{\sigma}\right)}} - \sum_{i=1}^n{\frac{x_i-\mu}{\sigma}} \right)}.
\end{equation}

Based on Rostami and Adam \cite{Rostami2013}, where eighteen pairs of priors were selected for Gumbel distribution and compared the posterior estimations by applying Metropolis-Hastings algorithm, in this model we choose the combination of Gumbel and Rayleigh as the most productive pair of priors, namely,
\begin{equation} 
\begin{split}
\pi(\mu)&\propto \exp \left( -\exp \left(-\frac{\mu-\mu_0}{\sigma_0}\right) -\frac{\mu-\mu_0}{\sigma_0} \right), \\
\pi(\sigma)&\propto \sigma \text{exp} \left(-\frac{\sigma^2}{2\lambda_0^2}\right).
\end{split}
\end{equation}
So, the posterior distribution is
\begin{equation}
\pi(\mu, \sigma|\mathbf{x}) \propto \frac{1}{\sigma^{n-1}} \text{exp}\left( A -\exp \left(-\frac{\mu-\mu_0}{\sigma_0}\right)-\frac{\mu-\mu_0}{\sigma_0} -\frac{\sigma^2}{2\lambda_0^2}\right),
\end{equation}
and the marginal posterior distributions are given by
	\begin{equation}
\pi(\mu|\mathbf{x}) \propto \text{exp}\left( A -\exp \left(-\frac{\mu-\mu_0}{\sigma_0}\right) -\frac{\mu-\mu_0}{\sigma_0} \right),
\end{equation}
and
\begin{equation}
\pi(\sigma|\mathbf{x}) \propto \frac{1}{\sigma^{n-1}} \text{exp}\left( A -\frac{\sigma^2}{2\lambda_0^2}\right).
\end{equation}
Here, $A=- \sum_{i=1}^n{\exp{\left(-\dfrac{x_i-\mu}{\sigma}\right)}} - \sum_{i=1}^n{\dfrac{x_i-\mu}{\sigma}}$.

The Metropolis-Hasting algorithm (MH) is a Markov chain Monte Carlo (MCMC) method for
collecting a sequence of random samples from a probability distribution where direct sampling is difficult \cite{amin2015bayesian}.
In our context, the simple steps of MH algorithm can be summarized as follows:

\begin{enumerate}
	\item Choose initial values: $\mu^{(0)},\sigma^{(0)}$
	
	\item Given the chain is currently at $\mu^{(j)},\sigma^{(j)}$:
	\begin{itemize}
	\item Draw a candidate $\mu^{\text{can}}, \sigma^{\text{can}}$ for the next sample, by picking from the normal distribution for some suitable chosen variances $v_{\mu}$ and $v_{\sigma}$,
	 $$\mu^{\text{can}} \sim \mathcal{N}(\mu^{(j)},v_{\mu}) \text{ and } \sigma^{\text{can}} \sim \mathcal{N}(\sigma^{(j)},v_{\sigma})$$
	\item Accept $\mu^{\text{can}}$ with probability 
$$P_{\mu}=\text{min}\left\{1,\frac{\pi(\mu^{\text{can}}|\sigma^{(j)})}{\pi(\mu^{(j)}|\sigma^{(j)})}\right\};$$ 
where

\begin{eqnarray}
\log \left( \dfrac{\pi(\mu^{\text{can}}|\sigma^{(j)})}{\pi(\mu^{(j)}|\sigma^{(j)})} \right)&=&\dfrac{n}{\sigma}(\mu^{\text{can}}-\mu^{(j)})+\dfrac{\mu^{(j)}-\mu^{\text{can}}}{\sigma_0}+ \nonumber \\
 & &\exp{\left( -\dfrac{\mu^{(j)}-\mu_0}{\sigma_0}\right)}-\exp{\left(-\dfrac{\mu^{\text{can}}-\mu_0}{\sigma_0}\right)}+\nonumber \\
 & &\sum_{i=1}^{n} \left( \exp{\left(- \dfrac{x_i-\mu^{(j)}}{\sigma^{(j)}}\right)} - \exp{\left(- \dfrac{x_i-\mu^{\text{can}}}{\sigma^{(j)}}\right)}  \right) \nonumber \\
\end{eqnarray}

		\item Accept $\sigma^{\text{can}}$ with probability 
$$P_{\sigma}=\text{min}\left\{1,\frac{\pi(\sigma^{\text{can}}|\mu^{(j)})}{\pi(\sigma^{(j)}|\mu^{(j)})}\right\}$$
where

\begin{eqnarray}
\log \left( \dfrac{\pi(\sigma^{\text{can}}|\mu^{(j)})}{\pi(\sigma^{(j)}|\mu^{(j)})} \right)&=&(1-n)(\log{\sigma^{\text{can}}}-\log{\sigma^{(j)}})+
\dfrac{\left( \sigma^{(j)}\right)^2-\left(\sigma^{\text{can}}\right)^2}{2\lambda_0^2}+ \nonumber \\
 & & \sum_{i=1}^n {\left(  \left( x_i-\mu^{(j)} \right) \left( \frac{1}{\sigma^{(j)}} - \frac{1}{\sigma^{\text{can}}} \right) \right)} + \nonumber \\
 & &\sum_{i=1}^{n} \left( \exp{\left(- \dfrac{x_i-\mu^{(j)}}{\sigma^{(j)}}\right)} - \exp{\left(- \dfrac{x_i-\mu^{(j)}}{\sigma^{\text{can}}}\right)}  \right) \nonumber \\
\end{eqnarray}
This is implemented by drawing $u_{\mu}, u_{\sigma} \sim \mathcal{U}(0,1)$ and taking $\mu^{(j+1)}=\mu^{\text{can}}$ if and only if $u_{\mu}<P_{\mu}$ and $\sigma^{(j+1)}=\sigma^{\text{can}}$ if and only if $u_{\sigma}<P_{\sigma}$.

\end{itemize}	
	\item Iterate the former procedure.
	
\end{enumerate}

When performing the MH algorithm some aspects must be considered. First, samples are highly correlated, so most of the MH draws have to be dropped by setting a suitable number $d$ (usually determined by analyzing the autocorrelation between samples), consequently, the $n \times d^{th} $ draws are the only archived data. Second, the choice of the initial values could be far from the real one. As a result, a $\textit{burn-in}$ period is necessary to pick values in the convergence plateau.

\section{Regression analysis}\label{reg}
As aforementioned, the aim of this paper is to find a relationship between the parameters of the Gumbel distribution and those ones of the baseline distribution. The easiest way to do so is by running the MH iterative procedure for a big enough sequence of parameters belonging to the distribution which generate the maximum dataset. Convergence of the distribution function for the extreme values is guaranteed for high sample sizes. Hence, the larger number of data is, the more accurate the regression coefficients will be. Furthermore, following the analysis done by Gumbel \cite{Gumbel2012}, the relation between the parameters will depend on the length of each sample or block ($k$) available to carry out the maximum selection and on the number of data ($n$), of course. 
\subsection{Simulation study}
One of the most distinctive features of the Gumbel distribution is its ability to represent the limit distribution for the maximum values of a wide range of baseline distributions. normal and exponential distribution fall within this range and provide mono and multi-parametric performances examples since they depend on the rate $\lambda$, and the mean, $\mu_N$, and standard deviation, $\sigma_N$, parameters, respectively. For the sake of convenience, a 90 sample size data for each distribution is selected in order to adapt the results to seasonal studies\footnote{Note that the sample size is a not alterable variable in all the studies and algorithms. Changes to this quantity will suppose a convergence to a Gumbel distribution with different parameters and, as a consequence,  regression coefficients will change.}.

\begin{figure}[H]
	\centering
	\subfloat{
		\label{fig:10m}
		\includegraphics[width=0.5\textwidth]{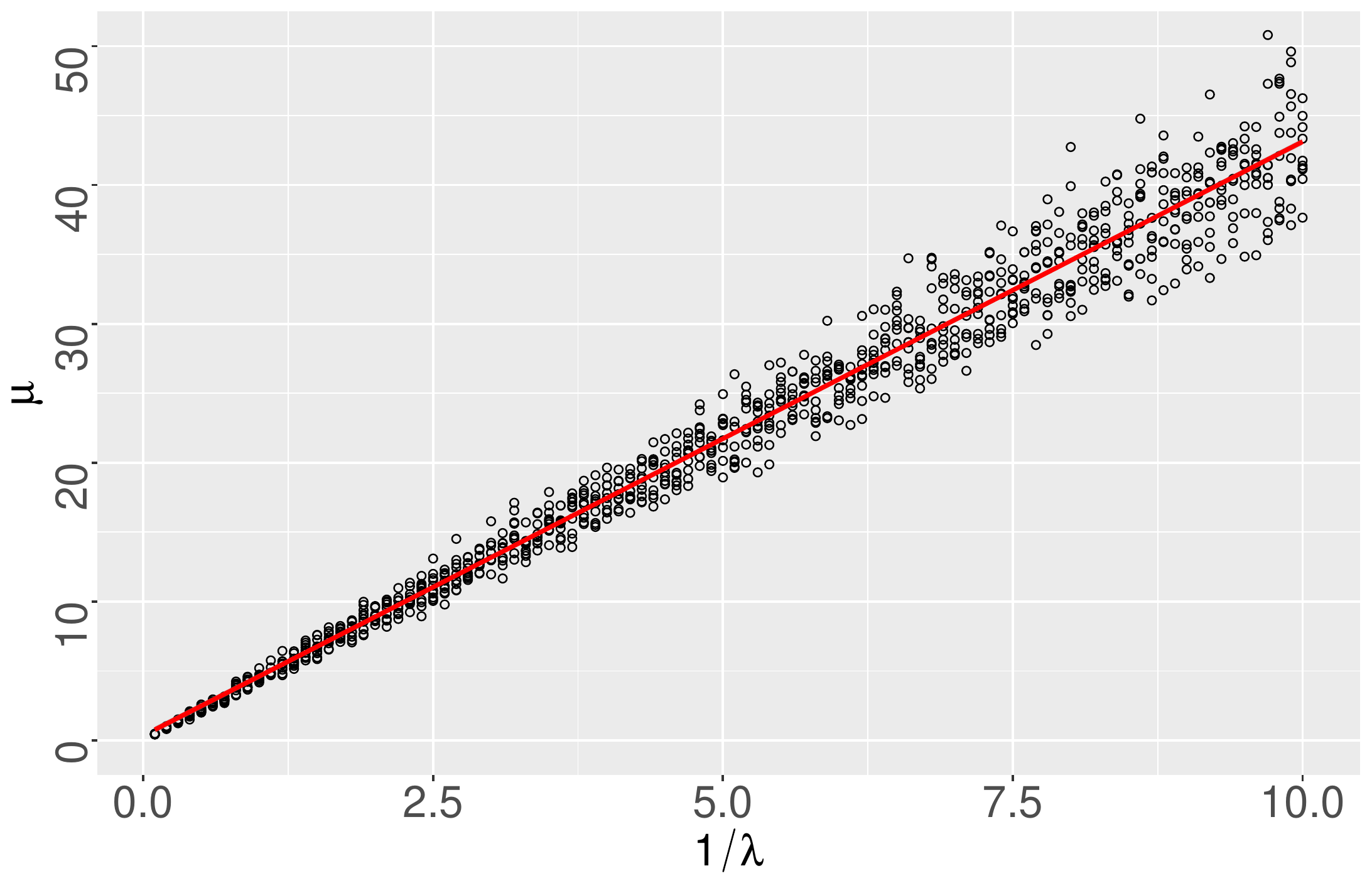}}
	\subfloat{
		\label{fig:1000m}
		\includegraphics[width=0.5\textwidth]{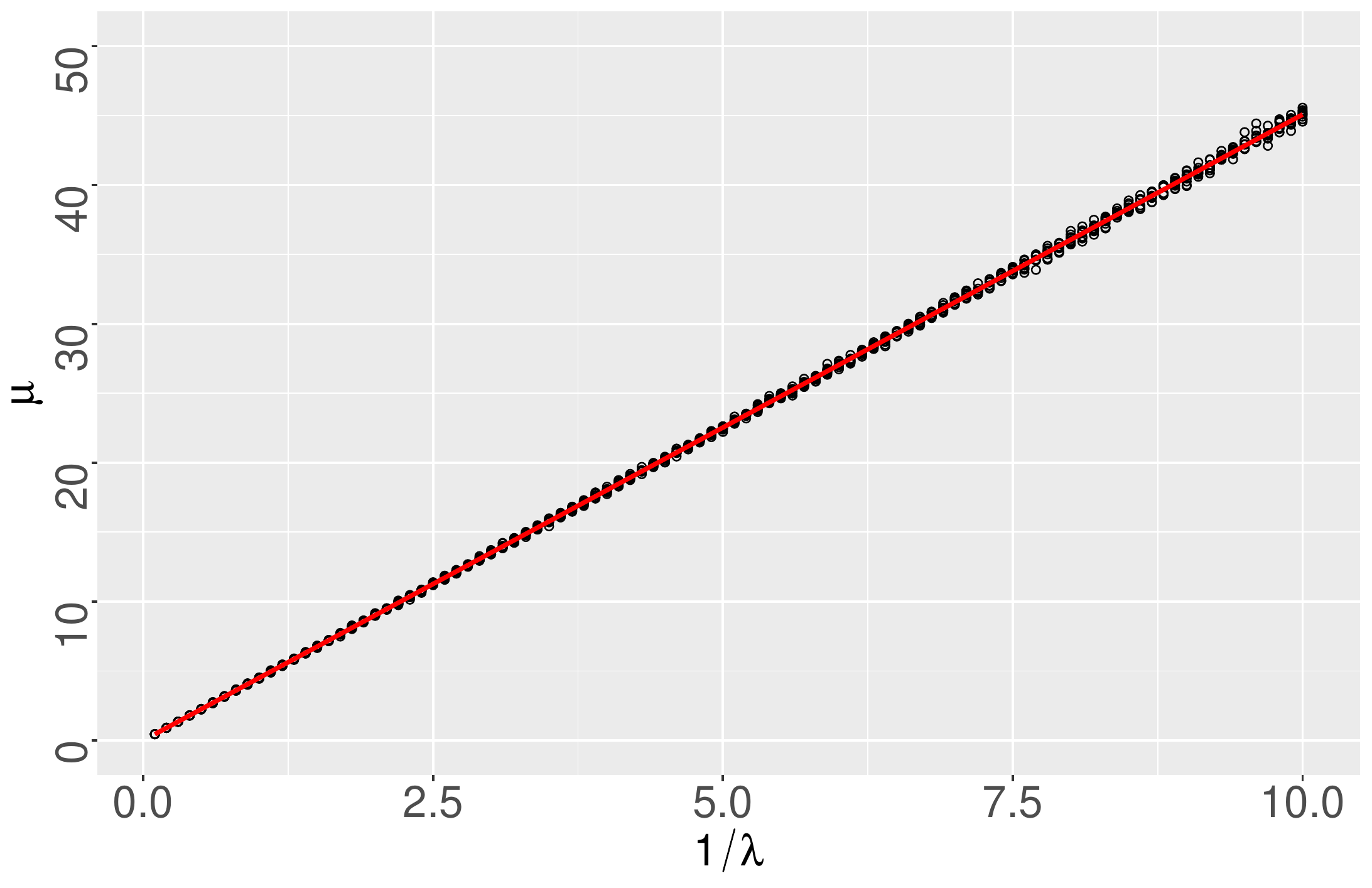}}\\
	
	\subfloat{
		\label{fig:10s}
		\includegraphics[width=0.5\textwidth]{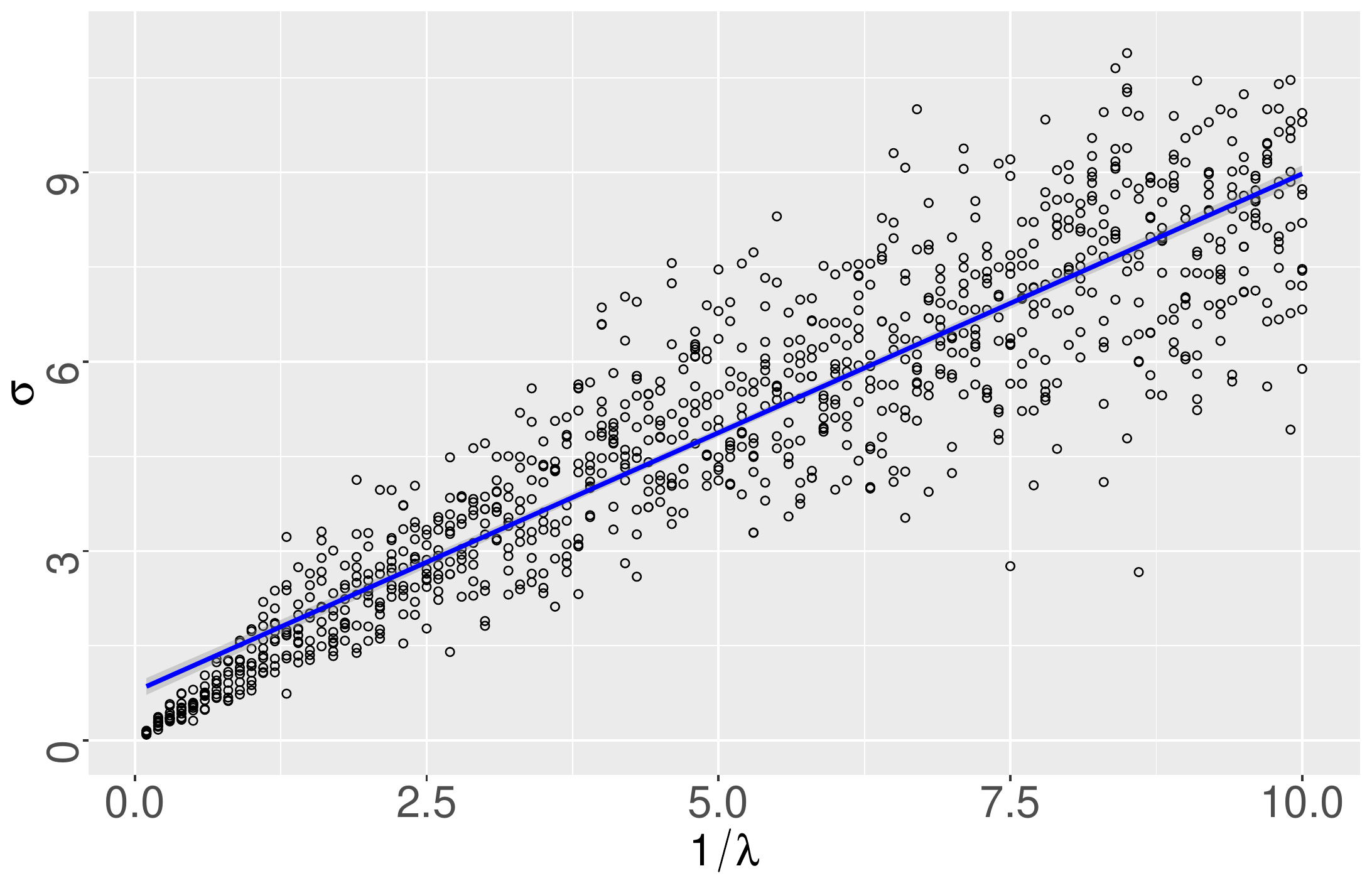}}
	\subfloat{
		\label{fig:1000s}
		\includegraphics[width=0.5\textwidth]{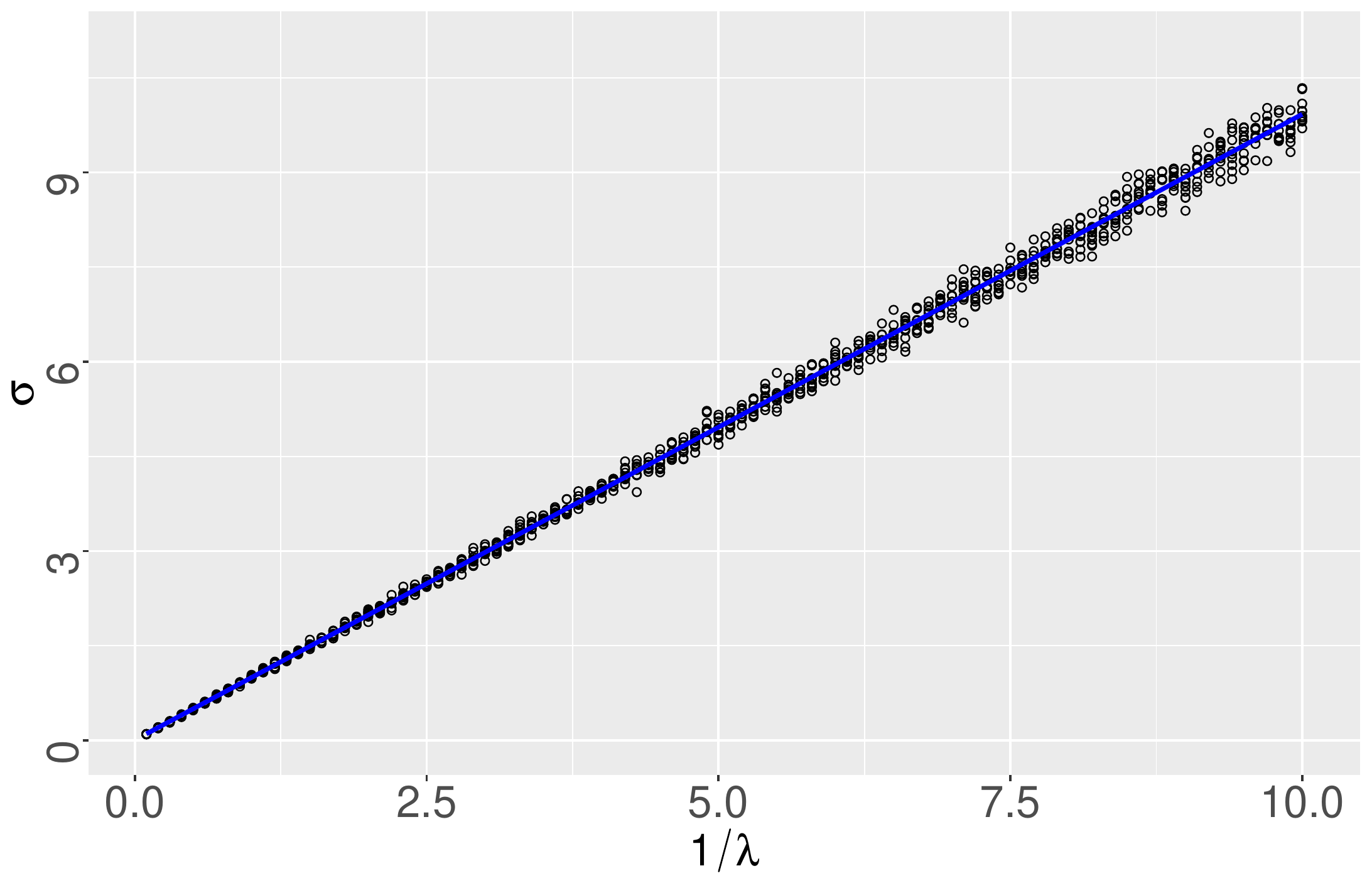}}
	
	\caption{Relation between the parameter of the exponential distribution (with values for $1/\lambda$ from 0.1 to 10,  by 0.1 regular increments) and $\mu$ (upper figures) and $\sigma$ (lower figures) parameters of the Gumbel distribution from 10 replicates of $n=10$ (left) and $n=1000$ (right) blocks of size $k=90$.}
	\label{fig:exps}
\end{figure}

\begin{figure}[H]
	\centering
	\subfloat{
		\label{fig:10nm}
		\includegraphics[width=0.5\textwidth]{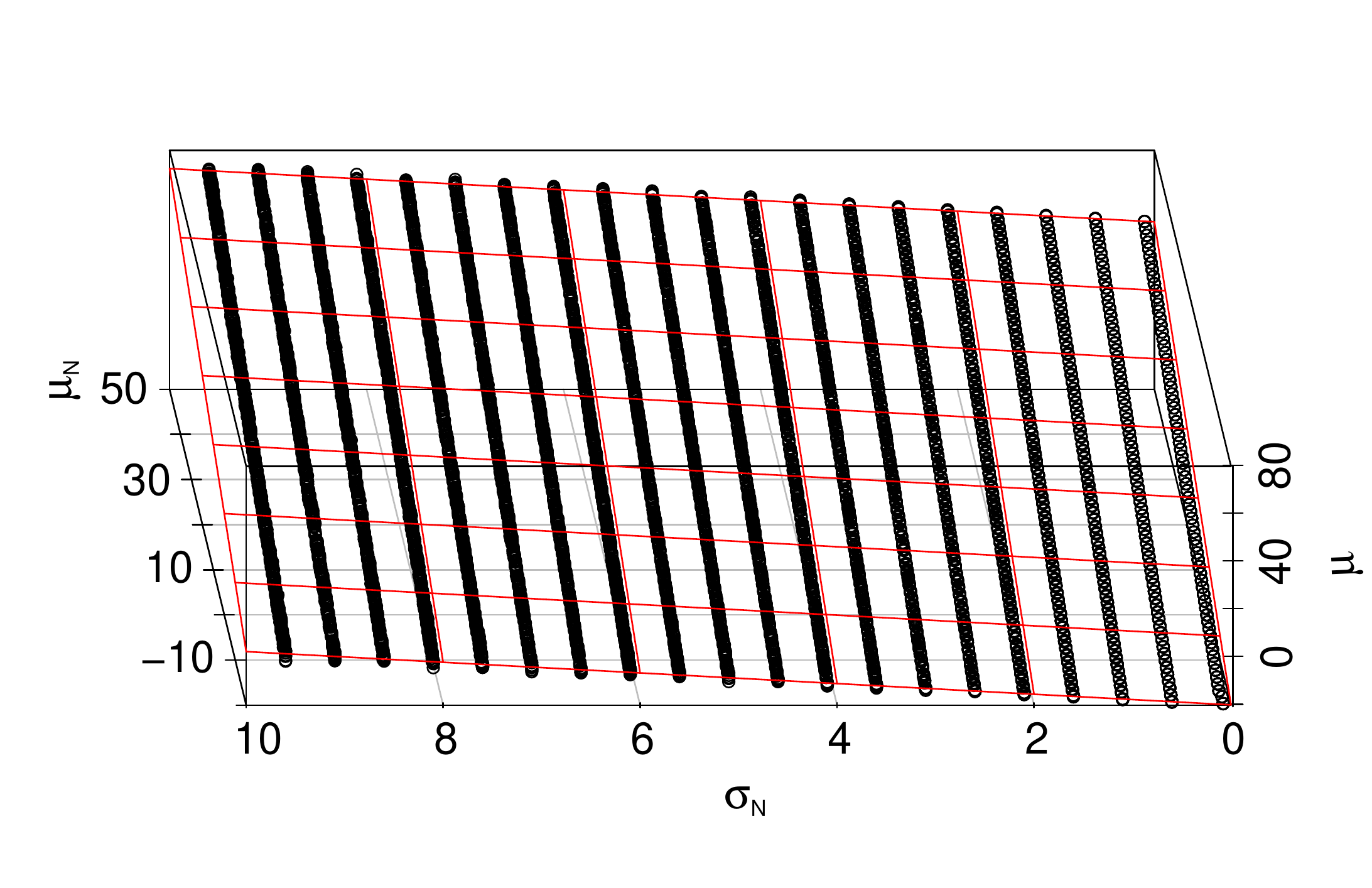}}
	\subfloat{
		\label{fig:1000nm}
		\includegraphics[width=0.5\textwidth]{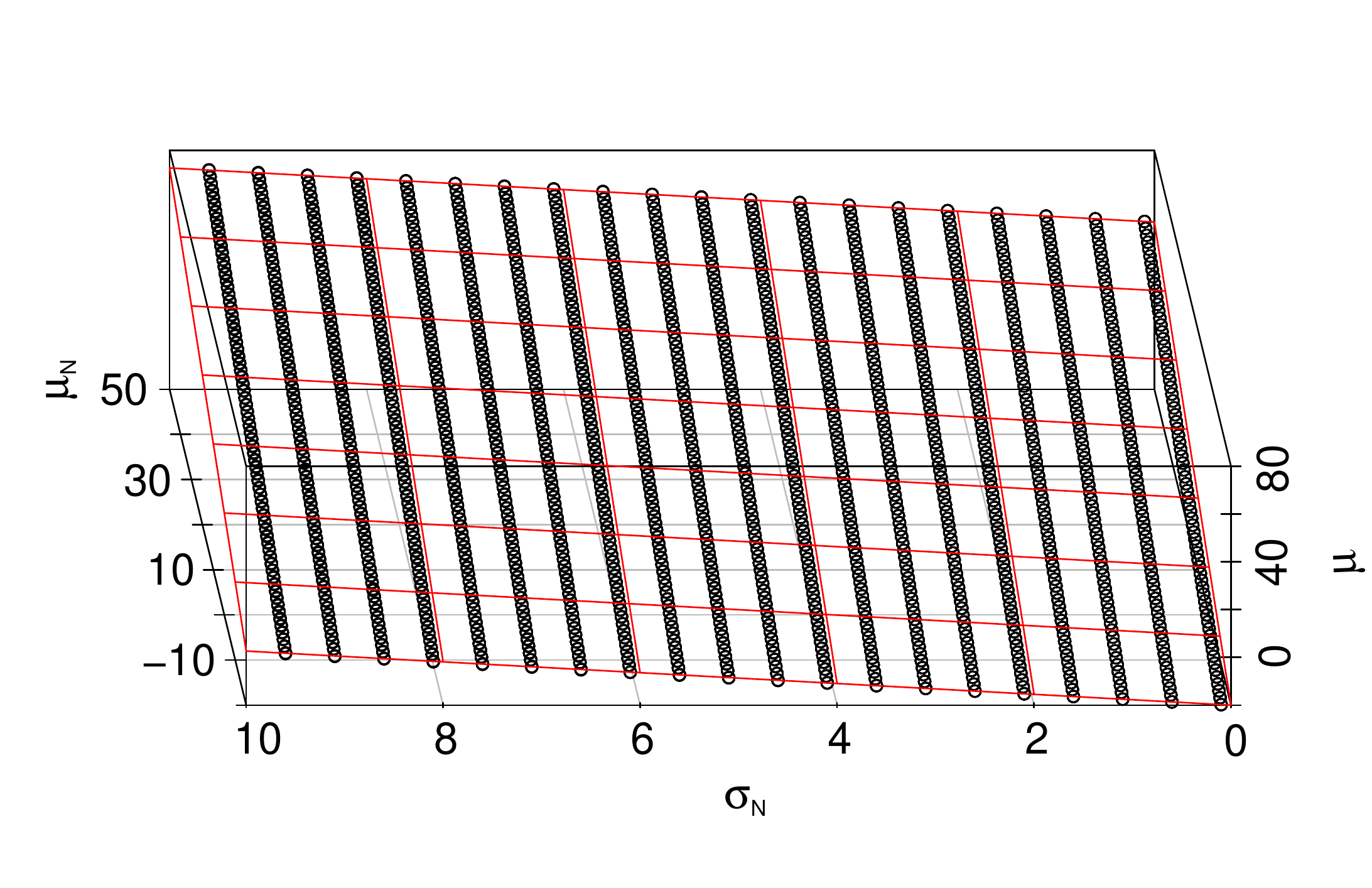}}\\
	
	\subfloat{
		\label{fig:10ns}
		\includegraphics[width=0.5\textwidth]{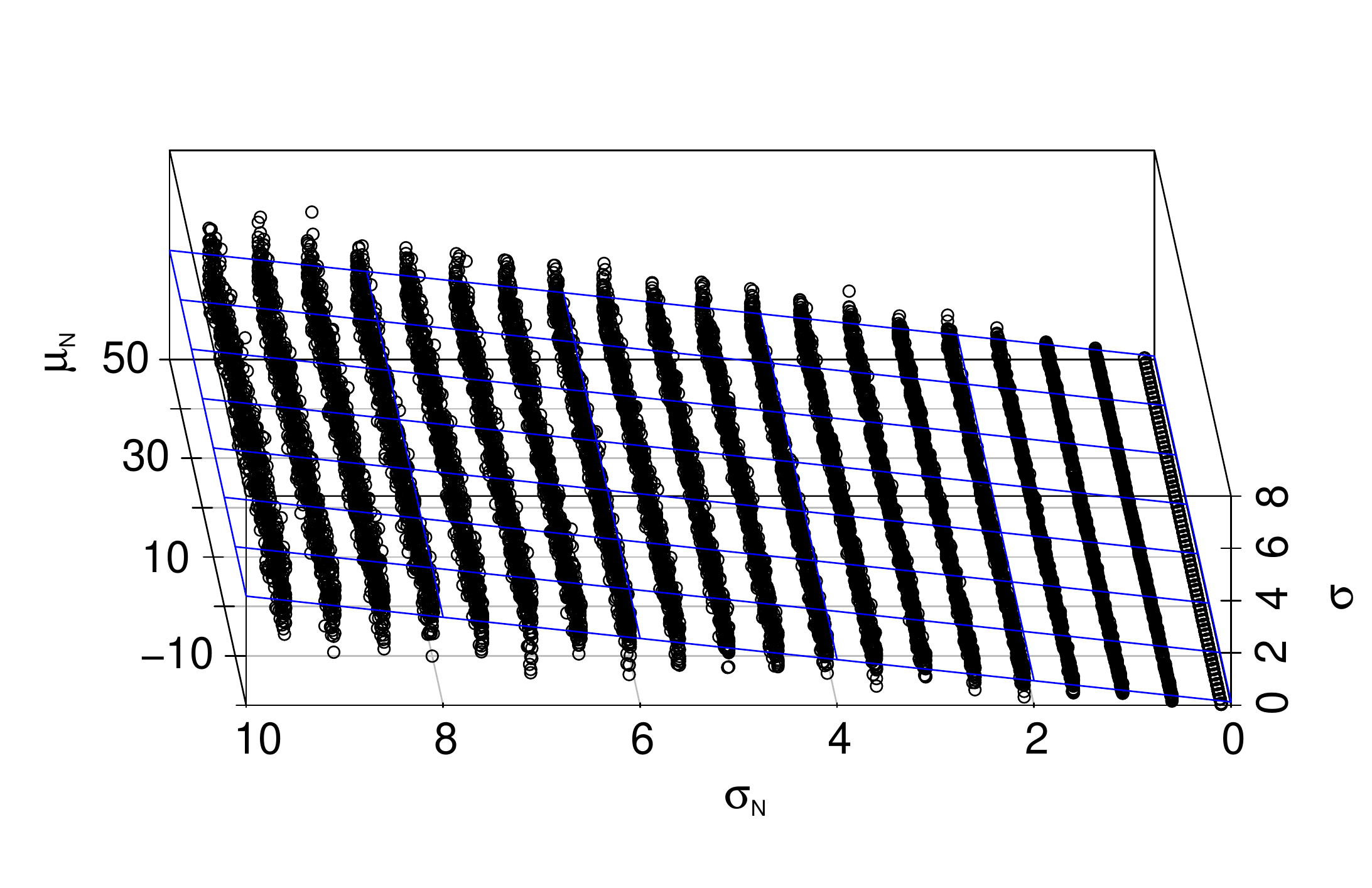}}
	\subfloat{
		\label{fig:1000ns}
		\includegraphics[width=0.5\textwidth]{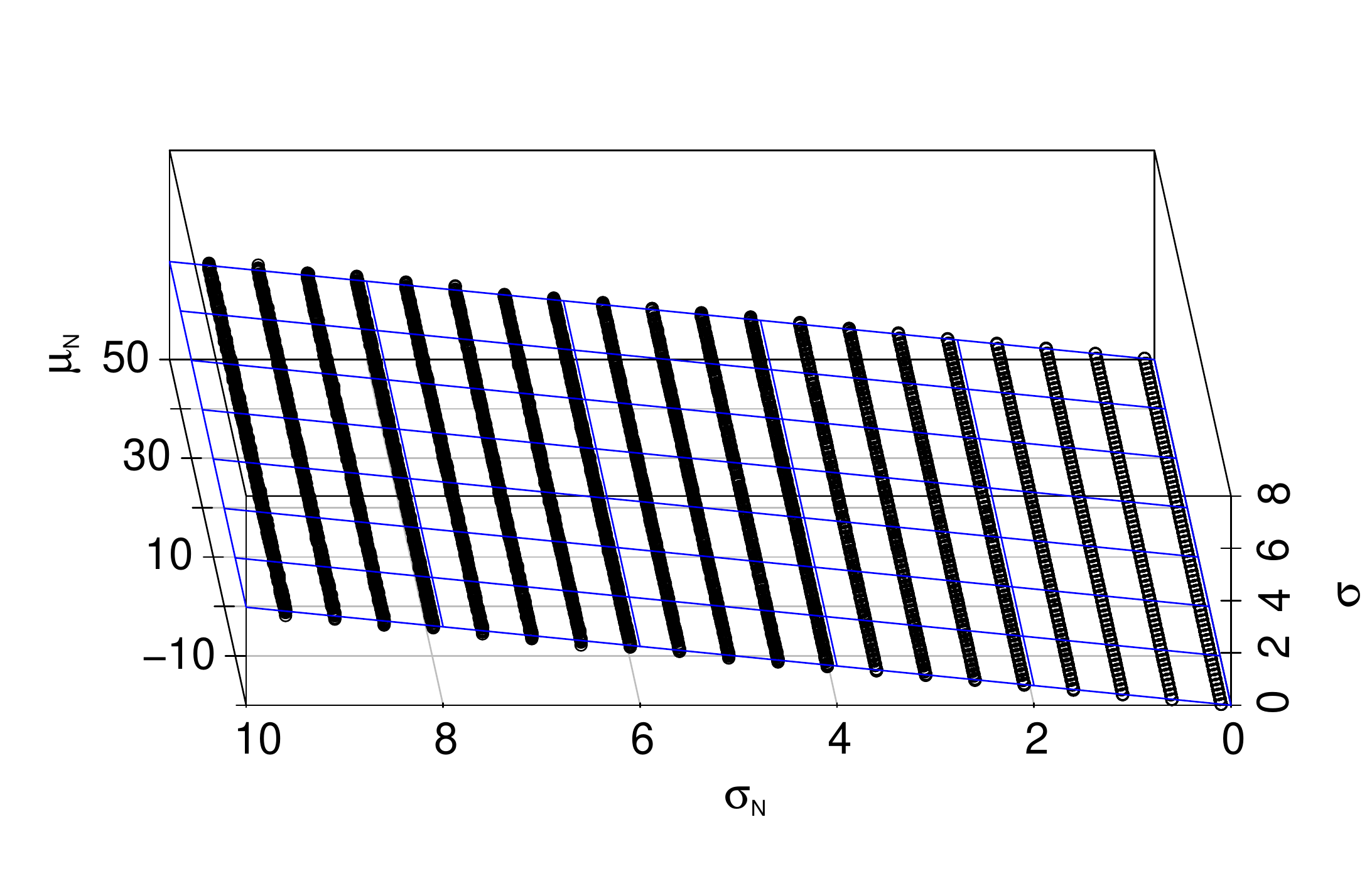}}
	
	\caption{Relation between the mean $\mu_N$ (values from -20 to 50,  by 1 regular increments) and standard deviation $\sigma_N$ (values from 0 to 10,  by 0.5 regular increments) parameters of the normal distribution and $\mu$ (upper figures) and $\sigma$ (lower figures) parameters of the Gumbel distribution from 10 replicates of $n=10$ (left) or $n=1000$ (right) samples of size $k=90$.} 	
	\label{fig:normal}
\end{figure}

Figures \ref{fig:exps} and \ref{fig:normal} illustrate the results from simulations for exponential and normal distribution, respectively. These plots consist in 10 replicate simulation results per parameter value with the aim of reducing the estimation uncertainty. As we can observe from the graphics, the existence of a clear relationship for the parameters is apparent. Moreover, their connections are not whichever but a linear one which will definitely simplify the priors implementation in successive algorithms. In addition, while increasing the number of samples, data dispersion substantially decreases. In this sense, posterior studies will focus on the use of information coming from data with high sample numbers. Regression results are summarized on Table \ref{tab:tabla1}. 

\begin{table}[h!]
	
	\centering
		\caption{
			Summarized samples from the posterior distribution of a linear regression to model the relationship between the Gumbel distribution parameters and exponential or normal baseline distribution parameters, obtained using the function \texttt{MCMCregress} by the R package \textit{MCMCpack} \cite{Martin2011}. Data frame is the table compound by $n=1000$ samples of size $k=90$.}
	\resizebox{\textwidth}{!}{
		
		\begin{tabular}{||c|c|c|c|c||}
			\hline
			\hline
			
			\multicolumn{5}{||c||}{Exponential Distribution}                                          \\
			\hline
			Parameter of the Gumbel Distribution&\multicolumn{2}{|c|}{  $\mu$}         & \multicolumn{2}{|c||}{  $\sigma$}       \\
			\hline
			Summarized posterior sample              & Mean      & SD        & Mean       & SD        \\
			\hline
			Intercept                           & -0.004409 & 0.012152  & 0.004305   & 0.0086136 \\
			\hline
			$1/\lambda$                                    & 4.507745  & 0.002109  & 0.991639   & 0.0014953 \\
			\hline
			\multicolumn{5}{||c||}{Normal Distribution}                                               \\
			\hline
			Parameter of the Gumbel Distribution&\multicolumn{2}{|c|}{  $\mu$}         & \multicolumn{2}{|c||}{  $\sigma$}       \\
			\hline
			Summarized posterior sample             & Mean      & SD        & Mean       & SD        \\
			\hline
			Intercept                            & 0.0001740 & 1.234e-03 & -4.909e-04 & 9.153e-04 \\
			\hline
			$\mu_N$                                    & 0.999978  & 2.874e-05 & -1.037e-05 & 2.132e-05 \\
			\hline
			$\sigma_N$                                    & 2.264677  & 2.062e-04 & 3.746e-01  & 1.530e-04\\
			\hline
			\hline
			
		\end{tabular}
	}

	\label{tab:tabla1}
\end{table}  
Outcomes given in Table 1 show a clear relationship between the parameters of the baseline distributions and the GEV ones. This information will be used in posterior studies to obtain informative priors based on data. Moreover, these priors can be used to generate candidates on the MH algorithm with the aim to accelerate convergence.

Although in this work we only display the results obtained for the normal and exponential distributions, this strategy could be easily adapted to other distributions.

\section{Regression Metropolis-Hasgtings method}\label{GM}
In the previous section, it has been established a relation between the parameters of the two most widely used baseline distributions and the Gumbel ones. Although the procedure can be applied to any baseline distribution we want, in this section it will be kept normal and exponential distributions as examples. 

The main objective of this section is to promote a generic model to import the highly informative regression coefficients into the Bayesian inference procedure. To do so, a normal prior distribution for each parameter of the Gumbel distribution is assumed in a manner that the mean and standard deviation are obtained from the regression coefficients
\begin{equation}\label{key}
\pi(\boldsymbol{\theta})\propto f(\boldsymbol{\alpha},\widetilde{\mathbf{x}}), 
\end{equation}
where $f$ is a function of the regression coefficients vector $\boldsymbol{\alpha}$ and the sample $\widetilde{\mathbf{x}}$. Then, $f$ can be used to estimate the parameters of the baseline distribution (noted by $\boldsymbol{\gamma}$) via a non-informative prior. Both regression and baseline parameters must be conveniently estimated. So, the full algorithm can be summarize as follows:

\begin{enumerate}
	\item Choose initial values, $\zeta^{(0)}=(\boldsymbol{\alpha}^{(0)},\boldsymbol{\gamma}^{(0)})$, where $\boldsymbol{\alpha}^{(0)}$ are previously calculated regression coefficients (Table \ref{tab:tabla1} showed examples for normal and exponential distributions) and $\boldsymbol{\gamma}^{(0)}$ are baseline parameters.
	
	\item Given the chain is currently at $\zeta^{(j)}=(\boldsymbol{\alpha}^{(j)}, \boldsymbol{\gamma}^{(j)})$:
\begin{itemize}
	\item Draw the candidates $\boldsymbol{\alpha}^{\text{can}}, \boldsymbol{\gamma}^{\text{can}}$ for the next sample by picking from the proper distribution. A normal distribution is used for $\boldsymbol{\alpha}^{\text{can}}$ with fixed mean and variance given by the regression coefficients, as well as a suitable chosen posterior distribution, calculated on the basis of a non-informative prior distribution and the entire dataset $\widetilde{\mathbf{x}}$, is used for $\boldsymbol{\gamma}^{\text{can}}$ (this distribution is specified in next subsection in the case of exponential and normal baseline distributions). 
	
	\item Due to the connection between $\boldsymbol{\gamma}$ and Gumbel parameters $\boldsymbol{\theta}$, $\boldsymbol{\alpha}^{\text{can}}$ and $\boldsymbol{\gamma}^{\text{can}}$ must be acepted or rejected simultaneously. As usual, this is implemented by drawing $u_{\boldsymbol{\theta}} \sim \mathcal{U}(0,1)$ and taking $\boldsymbol{\theta}^{(j+1)}=\boldsymbol{\theta}^{\text{can}}$ if and only if $u_{\boldsymbol{\theta}}<P_{\boldsymbol{\theta}}$.

	Here, 
	\begin{equation}
	P_{\boldsymbol{\theta}}=\text{min}\left\{1,\frac{\pi(\boldsymbol{\zeta}^{\text{can}}|\bold{x)}q(\boldsymbol{\zeta}^{(j)}|\boldsymbol{\zeta}^{\text{can}})}{\pi(\boldsymbol{\zeta}^{(j)}|\bold{x)}q(\boldsymbol{\zeta}^{\text{can}}|\boldsymbol{\zeta}^{(j)})}\right\},
	\end{equation}
	 where $\boldsymbol{\theta}$ can be $\mu$ or $\sigma$ and $\boldsymbol{\zeta}=(\boldsymbol{\alpha},\boldsymbol{\gamma})$. $\boldsymbol{\theta}$ is related with $\boldsymbol{\zeta}$ by
	\begin{equation}\label{key}
	\boldsymbol{\theta}=\boldsymbol{\alpha} (1,\boldsymbol{\gamma})'.
	\end{equation} 
	The conditional posterior distributions, $\pi(\boldsymbol{\theta}|\mathbf{x})$, for the parameters $\mu$ and $\sigma$ are obtained by ignoring
	all the terms that do not involved parameter $\mu$ and $\sigma$, respectively. For the sake of simplicity, and noting that the prior is highly informative, the candidates  are drawn from the prior distribution. Thus, $q(\boldsymbol{\zeta}^{(j)}|\boldsymbol{\zeta}^{\text{can}})=\pi(\boldsymbol{\zeta}^{(j)})$ and so,
	\begin{equation}
	P_{\boldsymbol{\theta}}=\text{min}\left\{1,\frac{L(\boldsymbol{\theta}^{\text{can}}|\mathbf{x})}{L(\boldsymbol{\theta}^{(j)}|\mathbf{x})}\right\}.
	\end{equation}
Here, $L(\boldsymbol{\theta}|\mathbf{x})$ is given by Eq. (\ref{likehood}), since it depends only on the maximum data. Despite this, the entire data $\widetilde{\mathbf{x}}$ is still used for generating draws of $\boldsymbol{\gamma}$.
	\end{itemize}
	 \item Iterate the former procedure.
	
\end{enumerate}
\subsection{Exponential and normal distribution examples}
Proceeding with the examples given in Section \ref{reg}, results of regression summarized in Table \ref{tab:tabla1} are used. To draw the candidates in the MH step, it is necessary to know the \textit{non-informative} posterior distribution for the parameter $\lambda$ of the exponential and, the $\mu$ and $\sigma$ parameters of the normal that will be used as priors in the iterative procedure. 

With respect to the exponential case, it is assumed a non-informative $\Gamma(\alpha=1,\beta=1)$ prior for $\lambda$. Then, posterior distribution is (for a data size $m$)

\begin{equation}\label{prior1}
\pi(\lambda|\widetilde{\mathbf{x}})\propto \operatorname{e}^{-\lambda \left( \sum_{i=1}^m x_i +1 \right)} \lambda^{m} \propto \Gamma\left(m+1,\sum_{i=1}^m x_i+1\right).
\end{equation} 
Concerning the normal distribution, it is assumed the following prior distribution for $\mu$ and $\sigma$
\begin{equation}
\begin{split}
\mu|\sigma&\sim \mathcal{N}(\mu_0=0,\sigma_0=1)\\
\sigma&\sim\Gamma(\alpha=1,\beta=1).
\end{split}
\end{equation}
Then , the posterior is 
\begin{equation}\label{prior2}
\begin{split}
\pi(\mu|\sigma,\widetilde{\mathbf{x}})&\propto\mathcal{N}\left(\frac{m}{m+1}\bar{x} ,m+1\right)\\
\pi(\sigma|\widetilde{\mathbf{x}})&\propto\Gamma\left(1+\frac{m}{2},1+\frac{1}{2}\sum_{i=1}^m(x_i-\bar{x})^2+\frac{m}{2(m+1)}\bar{x}^2\right).
\end{split}
\end{equation}
With prior distributions for the baseline parameters given by Eqs. (\ref{prior1}) and (\ref{prior2}), which will be used to draw the candidates in the MH steps, and likehood function (\ref{likehood}), generic method is ready to be implemented. It should be noted that some convergence and correlation tests must be done before making a definitive analysis. 
\section{Discussion}
The comparability of the two considered methods; the Standard-MH (SMH) algorithm and the Regression-MH (RMH) method introduced in Section \ref{GM}, is explored via simulations, involving various sample sizes and values of the parameter space.

For all cases, the first $1000$ iterations out of $10000$ had to be discarded since the plot does not converge to the stable values before $1000$th iteration for MH algorithm. Moreover, to avoid autocorrelation between successive candidates, it has been set down a thinning rate retaining each 60th element of the sequence \cite{Plummer2006}. 

\begin{figure}[H]
	\centering{
		\includegraphics[width=1\textwidth]{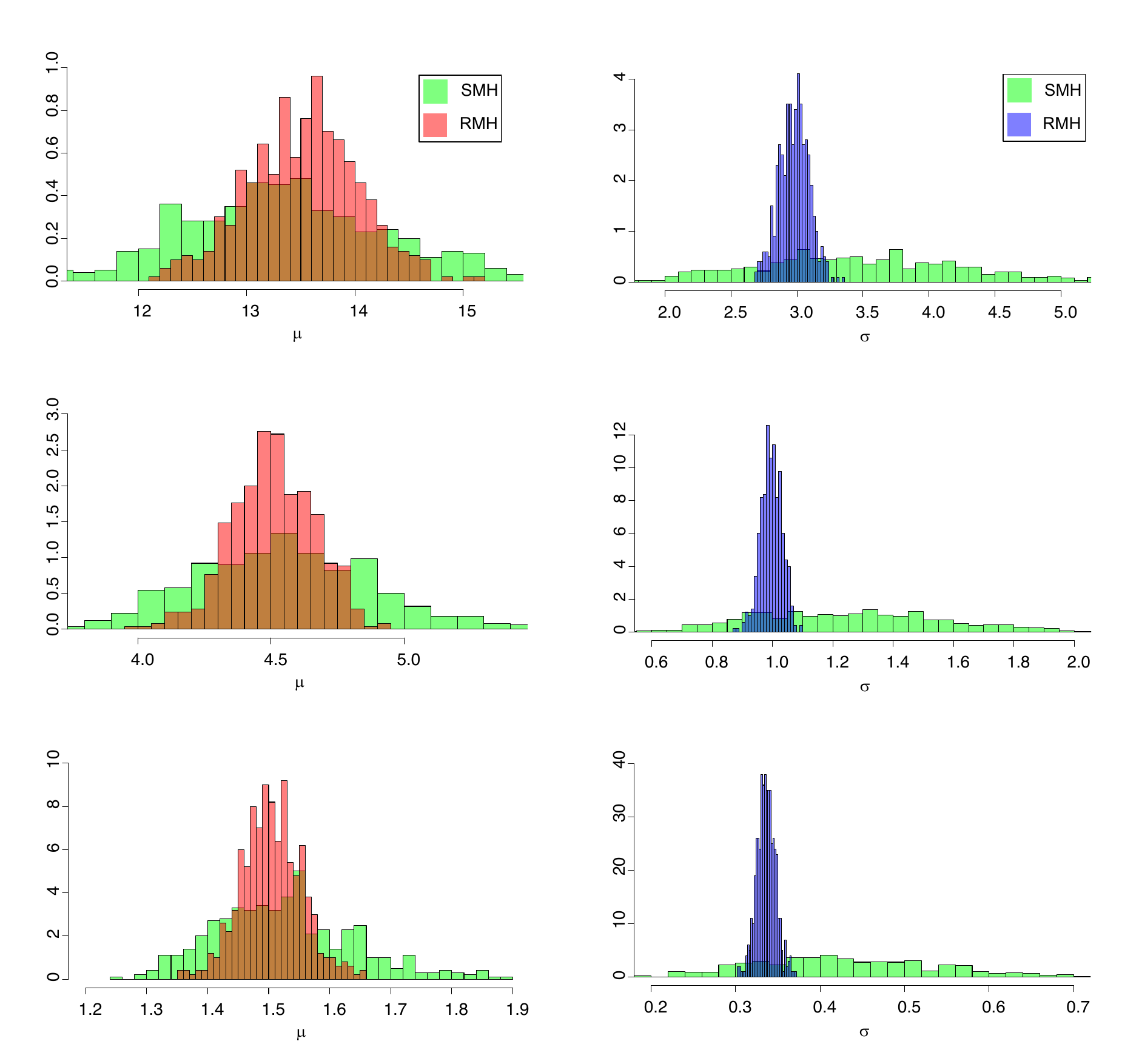}}
		
	\caption{ Histograms for estimated Gumbel parameters obtained using SMH and RMH methods, from 500 simulated samples with $k=90$, $n=10$ and baseline distribution exponential, with $\lambda=1/3$ (top), 1 (middle) and 3 (bottom).}
	\label{fig:hexp}
\end{figure}

\begin{figure}[H]
	\centering
	{
		\includegraphics[width=1\textwidth]{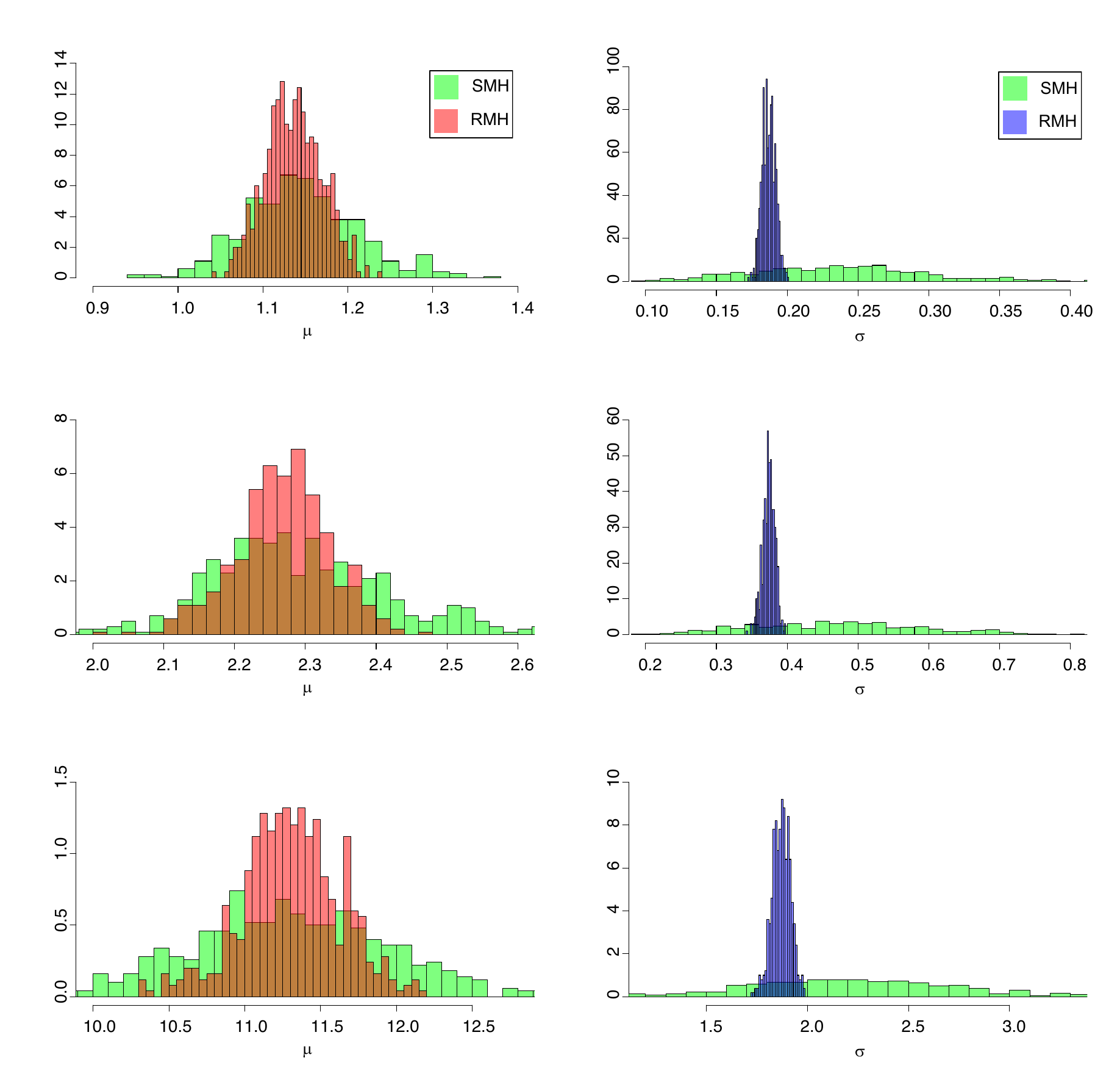}}
	
		\caption{Histograms for estimated Gumbel parameters obtained using SMH and RMH methods, from 500 simulated samples with $k=90$, $n=10$ and baseline distribution normal., with parameters $\mu_N=0$ and $\sigma_N=0.5$ (top), 1 (middle) and 5 (bottom).}
	\label{fig:hnormal}
\end{figure}

The main conclusion derived from this experiment is that an improvement has been achieved. Figures \ref{fig:hexp} and \ref{fig:hnormal} show the distribution of estimations obtained by the SMH and RMH methods for $\mu$ and $\sigma$, as particular examples of general behavior, which is more accused for large sample sizes. In the same way, the summary of the credible intervals for the estimated Gumbel parameters is shown in Table \ref{tab:table2}. Reduction in the credible interval width and, not less, enhacement in parameter location show the RMH results to adapt to very shortened data better than SMH does. Even outcomes show a distinct narrowing in the value range, this is most evident when looking at the scale parameter $\sigma$. The scale parameter can provide more significant information when talking about statistical inference. This means, to make better predictions of extremes events it is better to have a more accurate knowledge of the closeness of extreme values than knowing the central tendency of them because it will reduce the probability interval of finding these events.

\begin{table}[h!]
	\centering
		\caption{
			95\% Credible interval summary for exponential and normal baseline distribution using SMH and RMH methods, from 500 simulated samples with $k=90$, $n=10$, and baseline parameters $\lambda=1/3, 1, 3$, $\mu_N=0, 10$, and $\sigma_N=0.5, 1, 5$ respectively. 
	}
	\resizebox{\textwidth}{!}{
		
		\begin{tabular}{||c|c|c|c|c||}
			\hline
			\hline
			
			\multicolumn{5}{||c||}{Exponential Distribution}                                          \\
			\hline
			Parameter of the Gumbel Distribution&\multicolumn{2}{|c|}{  $\mu$}         & \multicolumn{2}{|c||}{  $\sigma$}       \\
			\hline
			Method            &SMH     & RMH       &SMH     & RMH        \\
			\hline
			$\lambda=1/3$                          &(11.575,15.272) & (12.473,14.520) & (2.027,5.007) & (2.749,3.199) \\
			$\lambda=1$                          &(3.952,5.256) & (4.167,4.786) & (0.711,1.899) & (0.922,1.058)\\
			$\lambda=3$                          &(1.333,1.791) & (1.406,1.609) & (0.245,0.672) & (0.3149,0.359) \\
			\hline
			\multicolumn{5}{||c||}{Normal Distribution}  \\
			\hline                                             
			Parameter of the Gumbel Distribution&\multicolumn{2}{|c|}{  $\mu$}         & \multicolumn{2}{|c||}{  $\sigma$}       \\
			\hline
			Method            &SMH     & RMH       &SMH     & RMH        \\       
			\hline
			$\mu_N=0, \sigma_N=0.5$               & (1.023,1.295) & (1.075,1.204) & (0.130,0.362)  & (0.179,0.196) \\
			$\mu_N=0, \sigma_N=1$               & (2.058,72.555) & (2.129,2.391) & (0.261,0.712) & (0.356,0.389) \\
			$\mu_N=0, \sigma_N=5$               &(10.096,12.663) & (10.594,11.912) & (1.297,3.232) & (1.776,1.953) \\
			\hline
			$\mu_N=10, \sigma_N=0.5$               &(11.031,11.277) & (11.087,11.216) & (0.127,0.379)  & (0.181,0.200) \\
			$\mu_N=10, \sigma_N=1$               & (12.076,12.540) & (12.145,12.404) & (0.275,0.716) & (0.358,0.392) \\
			$\mu_N=10, \sigma_N=5$               & (20.111,22.438) & (20.634,21.994) & (1.313,3.265) & (1.783,1.960) \\
			\hline
			\hline
			
		\end{tabular}
	}

	\label{tab:table2}
\end{table}


\section{An experimental data example}
With the aim of giving a practical application of the RMH model, the science of climatology will be considered. This science and its practical application have much to contribute to EVT. In particular, RMH algorithm will be used to estimate the Gumbel parameters that fit the peak temperatures in summer seasons (June-August). These seasons will be treated separately because of the time dependence of the normal distribution parameters that describe summer temperatures. For this reason, actual data consists in 90 maximum daily summer temperatures and the maximum value of each season. The dataset used in this paper consists of a set of daily temperature records registered in the city of C\'aceres (Extremadura, Spain) in the 1908-1915 period. 

Results are displayed on Table \ref{tab:table4}. It can be noted that credible intervals are extraordinary shorts, in view especially of the fact that only one maximum value per season have been used. However, this can be explained by the fact that not only one singular datum has been used, but all of the entire data. The capacities of the RMH model to adapt to very shortened data is the main advantage of this model with respect to those which target efforts on improving prior information of the maximum distribution.
\begin{table}[H]
	\centering
		\caption{
		Credible interval summary for the Gumbel parameters arising from a normally distributed annual temperature data.
	}
	\resizebox{\textwidth}{!}{
		
		\begin{tabular}{||c|c|c|c|c||}
			\hline
			\hline
			Year&\multicolumn{2}{|c|}{  1908}         & \multicolumn{2}{|c||}{  1909}       \\
			\hline
			Parameter of the Gumbel Distribution ($^{\circ}$C)            &$\mu$    & $\sigma$       &$\mu$     & $\sigma$        \\
			\hline
			95$\%$ Credible Interval                            &(42.026,42.052) &(1.847,1.851)   &(44.304,44.336)  & (2.164,2.168)   \\
			\hline                                             
			Year&\multicolumn{2}{|c|}{  1910}         & \multicolumn{2}{|c||}{  1911}       \\
			\hline
			Parameter of the Gumbel Distribution ($^{\circ}$C)            &$\mu$    & $\sigma$       &$\mu$     & $\sigma$        \\
			\hline
			95$\%$ Credible Interval                            &(42.429,42.456) & (1.866,1.870)  & (44.531,44.562)  &(2.117,2.121) \\
			\hline   
			Year&\multicolumn{2}{|c|}{  1912}         & \multicolumn{2}{|c||}{  1913}       \\
			\hline
			Parameter of the Gumbel Distribution ($^{\circ}$C)            &$\mu$    & $\sigma$       &$\mu$     & $\sigma$        \\
			\hline
			95$\%$ Credible Interval                            &(41.243,41.271) & (1.811,1.815)  & (43.414,43.441)  &(1.784,1.788) \\
			\hline
			Year&\multicolumn{2}{|c|}{  1914}         & \multicolumn{2}{|c||}{  1915}       \\
			\hline
			Parameter of the Gumbel Distribution ($^{\circ}$C)            &$\mu$    & $\sigma$       &$\mu$     & $\sigma$        \\
			\hline
			95$\%$ Credible Interval                            &(41.737,41.761) & (1.774,1.777)  & (42.847,42.872)  &(1.611,1.614) \\
			\hline 
			\hline
			
		\end{tabular}
	}

	\label{tab:table4}
\end{table}
\section{Concluding remarks}
In this paper, a new method to incorporate Bayesian methodology into extreme value analysis is studied. A Bayesian MH algorithm based on highly informative priors obtained by a well-defined connection between Gumbel and baseline parameters is presented. Simulations made show that the proposed method (RMH) can accurately estimate the Gumbel parameters and considerably shorten the credible interval width when compared with the standard MH algorithm. Experimental data fits show that the proposed model is also suitable when working with only one maximum value. It is found that RMH method has good performance and its usage is recommended in practice.

Although we only show the results obtained for the normal and exponential baseline distributions and extreme distribution Gumbel, the strategy can be extended to other distributions.

\section{Acknowledgments}
Thanks are due to the Spanish State Meteorological Agency (Agencia Estatal de Meteorolog\'ia: www.aemet.es) for providing the daily temperature time series used in this study. The present work has been supported by the Ministerio de Industria, Economía y Competitividad (Spain) through Grants Nos. MTM2014-56949-C3-3-R and MTM2017-86875-C3-2-R, partially financed by FEDER funds and by the Junta de Extremadura (Spain) through Grant No. IB16063.
\section*{References}

\bibliography{bibtex}

\begin{thebibliography}{10}
\expandafter\ifx\csname url\endcsname\relax
  \def\url#1{\texttt{#1}}\fi
\expandafter\ifx\csname urlprefix\endcsname\relax\def\urlprefix{URL }\fi
\expandafter\ifx\csname href\endcsname\relax
  \def\href#1#2{#2} \def\path#1{#1}\fi

\bibitem{nogaj2006amplitude}
M.~Nogaj, P.~Yiou, S.~Parey, F.~Malek, P.~Naveau, Amplitude and frequency of
  temperature extremes over the north atlantic region, Geophysical Research
  Letters 33~(10).
\newblock \href {http://dx.doi.org/10.1029/2005gl024251}
  {\path{doi:10.1029/2005gl024251}}.

\bibitem{Coelho2008}
C.~A.~S. Coelho, C.~A.~T. Ferro, D.~B. Stephenson, D.~J. Steinskog, Methods for
  exploring spatial and temporal variability of extreme events in climate data,
  Journal of Climate 21~(10) (2008) 2072--2092.
\newblock \href {http://dx.doi.org/10.1175/2007jcli1781.1}
  {\path{doi:10.1175/2007jcli1781.1}}.

\bibitem{Acero2017}
F.~J. Acero, M.~I. Fern{\'a}ndez-Fern{\'a}ndez, V.~M.~S. Carrasco, S.~Parey,
  T.~T.~H. Hoang, D.~Dacunha-Castelle, J.~A. Garc{\'i}a, Changes in heat wave
  characteristics over extremadura (sw spain), Theoretical and Applied
  Climatology (2017) 1--13\href {http://dx.doi.org/10.1007/s00704-017-2210-x}
  {\path{doi:10.1007/s00704-017-2210-x}}.

\bibitem{Garcia2007}
J.~Garc{\'{\i}}a, M.~C. Gallego, A.~Serrano, J.~Vaquero, Trends in
  block-seasonal extreme rainfall over the iberian peninsula in the second half
  of the twentieth century, Journal of Climate 20~(1) (2007) 113--130.
\newblock \href {http://dx.doi.org/10.1175/jcli3995.1}
  {\path{doi:10.1175/jcli3995.1}}.

\bibitem{Re2009}
M.~Re, V.~R. Barros, Extreme rainfalls in {SE} south america, Climatic Change
  96~(1-2) (2009) 119--136.
\newblock \href {http://dx.doi.org/10.1007/s10584-009-9619-x}
  {\path{doi:10.1007/s10584-009-9619-x}}.

\bibitem{Acero2011}
F.~J. Acero, J.~A. Garc{\'{\i}}a, M.~C. Gallego, Peaks-over-threshold study of
  trends in extreme rainfall over the iberian peninsula, Journal of Climate
  24~(4) (2011) 1089--1105.
\newblock \href {http://dx.doi.org/10.1175/2010jcli3627.1}
  {\path{doi:10.1175/2010jcli3627.1}}.

\bibitem{Acero2011a}
F.~J. Acero, M.~C. Gallego, J.~A. Garc{\'{\i}}a, Multi-day rainfall trends over
  the iberian peninsula, Theoretical and Applied Climatology 108~(3-4) (2011)
  411--423.
\newblock \href {http://dx.doi.org/10.1007/s00704-011-0534-5}
  {\path{doi:10.1007/s00704-011-0534-5}}.

\bibitem{Acero2017b}
F.~J. Acero, S.~Parey, T.~T.~H. Hoang, D.~Dacunha-Castelle, J.~A.
  Garc{\'{\i}}a, M.~C. Gallego, Non-stationary future return levels for extreme
  rainfall over extremadura (southwestern iberian peninsula), Hydrological
  Sciences Journal 62~(9) (2017) 1394--1411.
\newblock \href {http://dx.doi.org/10.1080/02626667.2017.1328559}
  {\path{doi:10.1080/02626667.2017.1328559}}.

\bibitem{Wi2015}
S.~Wi, J.~B. Vald{\'{e}}s, S.~Steinschneider, T.-W. Kim, Non-stationary
  frequency analysis of extreme precipitation in south korea using
  peaks-over-threshold and annual maxima, Stochastic Environmental Research and
  Risk Assessment 30~(2) (2015) 583--606.
\newblock \href {http://dx.doi.org/10.1007/s00477-015-1180-8}
  {\path{doi:10.1007/s00477-015-1180-8}}.

\bibitem{Ramos2007}
A.~A. Ramos, Extreme value theory and the solar cycle, Astronomy {\&}
  Astrophysics 472~(1) (2007) 293--298.
\newblock \href {http://dx.doi.org/10.1051/0004-6361:20077574}
  {\path{doi:10.1051/0004-6361:20077574}}.

\bibitem{Acero2017c}
F.~J. Acero, V.~M.~S. Carrasco, M.~C. Gallego, J.~A. Garc{\'{\i}}a, J.~M.
  Vaquero, Extreme value theory and the new sunspot number series, The
  Astrophysical Journal 839~(2) (2017) 98.
\newblock \href {http://dx.doi.org/10.3847/1538-4357/aa69bc}
  {\path{doi:10.3847/1538-4357/aa69bc}}.

\bibitem{Acero2018a}
F.~J. Acero, M.~C. Gallego, J.~A. Garc{\'{\i}}a, I.~G. Usoskin, J.~M. Vaquero,
  Extreme value theory applied to the millennial sunspot number series, The
  Astrophysical Journal 853~(1) (2018) 80.
\newblock \href {http://dx.doi.org/10.3847/1538-4357/aaa406}
  {\path{doi:10.3847/1538-4357/aaa406}}.

\bibitem{Castillo2004}
E.~Castillo, A.~S. Hadi, N.~Balakrishnan, J.~M. Sarabia, Extreme Value and
  Related Models with Applications in Engineering and Science,
  Wiley-Interscience, 2004.

\bibitem{Gumbel2012}
E.~J. Gumbel, Statistics of Extremes (Dover Books on Mathematics), Dover
  Publications, 2012.

\bibitem{castillo1987estadistica}
E.~Castillo~Ron, Estad{\'\i}stica de valores extremos. distribuciones
  asint{\'o}ticas, Estad{\'\i}stica espa{\~n}ola~(116) (1987) 5--35.

\bibitem{Smith1987a}
R.~L. Smith, J.~C. Naylor, A comparison of maximum likelihood and bayesian
  estimators for the three- parameter weibull distribution, Applied Statistics
  36~(3) (1987) 358.
\newblock \href {http://dx.doi.org/10.2307/2347795}
  {\path{doi:10.2307/2347795}}.

\bibitem{Coles2003a}
S.~Coles, L.~R. Pericchi, S.~Sisson, A fully probabilistic approach to extreme
  rainfall modeling, Journal of Hydrology 273~(1-4) (2003) 35--50.
\newblock \href {http://dx.doi.org/10.1016/s0022-1694(02)00353-0}
  {\path{doi:10.1016/s0022-1694(02)00353-0}}.

\bibitem{Bernardo1994}
J.~M. Bernardo, A.~F.~M. Smith (Eds.), Bayesian Theory, John Wiley {\&} Sons,
  Inc., 1994.
\newblock \href {http://dx.doi.org/10.1002/9780470316870}
  {\path{doi:10.1002/9780470316870}}.

\bibitem{Kotz2000}
S.~Kotz, S.~Nadarajah, Extreme Value Distributions: Theory and Applications,
  ICP, 2000.

\bibitem{Coles1996}
S.~G. Coles, J.~A. Tawn, A bayesian analysis of extreme rainfall data, Applied
  Statistics 45~(4) (1996) 463.
\newblock \href {http://dx.doi.org/10.2307/2986068}
  {\path{doi:10.2307/2986068}}.

\bibitem{Rostami2013}
M.~Rostami, M.~B. Adam, Analyses of prior selections for gumbel distribution,
  Matematika 29 (2013) 95--107.
\newblock \href {http://dx.doi.org/10.11113/matematika.v29.n.582}
  {\path{doi:10.11113/matematika.v29.n.582}}.

\bibitem{Chen2000}
M.-H. Chen, Q.-M. Shao, J.~G. Ibrahim, Monte Carlo Methods in Bayesian
  Computation, Springer New York, 2000.
\newblock \href {http://dx.doi.org/10.1007/978-1-4612-1276-8}
  {\path{doi:10.1007/978-1-4612-1276-8}}.

\bibitem{Vidal2013}
I.~Vidal, A bayesian analysis of the gumbel distribution: an application to
  extreme rainfall data, Stochastic Environmental Research and Risk Assessment
  28~(3) (2013) 571--582.
\newblock \href {http://dx.doi.org/10.1007/s00477-013-0773-3}
  {\path{doi:10.1007/s00477-013-0773-3}}.

\bibitem{Lye1990}
L.~M. Lye, Bayes estimate of the probability of exceedance of annual floods,
  Stochastic Hydrology and Hydraulics 4~(1) (1990) 55--64.
\newblock \href {http://dx.doi.org/10.1007/bf01547732}
  {\path{doi:10.1007/bf01547732}}.

\bibitem{Rostami2013a}
M.~Rostami, M.~B. Adam, N.~A. Ibrahim, M.~H. Yahya, Slice sampling technique in
  bayesian extreme of gold price modelling, in: AIP Conference Proceedings,
  Vol. 1557, AIP, 2013, pp. 473--477.
\newblock \href {http://dx.doi.org/10.1063/1.4823959}
  {\path{doi:10.1063/1.4823959}}.

\bibitem{1996}
S.~R. W.R.~Gilks, D.~Spiegelhalter, Markov Chain Monte Carlo in Practice
  (Chapman \& Hall/CRC Interdisciplinary Statistics), Chapman and Hall/CRC,
  1996.

\bibitem{amin2015bayesian}
N.~A.~M. Amin, M.~B. Adam, N.~A. Ibrahim, Bayesian inference using multiple-try
  metropolis hastings scheme for the efficiency of estimating gumbel
  distribution parameters, Matematika 31~(1) (2015) 25--36.
\newblock \href {http://dx.doi.org/10.11113/matematika.v31.n1.743}
  {\path{doi:10.11113/matematika.v31.n1.743}}.

\bibitem{Martin2011}
A.~D. Martin, K.~M. Quinn, J.~H. Park,
  \href{http://www.jstatsoft.org/v42/i09/}{{MCMCpack}: Markov chain monte carlo
  in {R}}, Journal of Statistical Software 42~(9) (2011) 22.
\newline\urlprefix\url{http://www.jstatsoft.org/v42/i09/}

\bibitem{Plummer2006}
M.~Plummer, N.~Best, K.~Cowles, K.~Vines,
  \href{https://journal.r-project.org/archive/}{Coda: Convergence diagnosis and
  output analysis for mcmc}, R News 6~(1) (2006) 7--11.
\newline\urlprefix\url{https://journal.r-project.org/archive/}

\end{thebibliography}

\end{document}